\documentclass[aps,prl,twocolumn,superscriptaddress]{revtex4}
\usepackage{graphicx}
\usepackage{siunitx}
\usepackage{bm}
\usepackage[caption=false]{subfig}
\usepackage{mathtools}
\usepackage{multirow}
\usepackage{amssymb}
\usepackage{braket}
\usepackage{amsmath}
\usepackage{placeins}
\usepackage{cancel}
\usepackage{xcolor}
\usepackage{spreadtab}

\begin{document}

\title{Route to high hole mobility in GaN via reversal of crystal-field splitting}

\author{Samuel Ponc\'e}
\affiliation{%
Department of Materials, University of Oxford, Parks Road, Oxford, OX1 3PH, UK
}%
\author{Debdeep Jena}
\affiliation{%
School of Electrical and Computer Engineering, Cornell University, Ithaca, New York 14853, USA
}
\affiliation{%
Department of Material Science and Engineering, Cornell University, Ithaca, New York 14853, USA
}%
\author{Feliciano Giustino}
\email{feliciano.giustino@materials.ox.ac.uk}
\affiliation{%
Department of Materials, University of Oxford, Parks Road, Oxford, OX1 3PH, UK
}%
\affiliation{%
Department of Material Science and Engineering, Cornell University, Ithaca, New York 14853, USA
}%

\date{\today}

\begin{abstract}
A fundamental obstacle toward the realization of GaN p-channel transistors is its low hole mobility. 
Here we investigate the intrinsic phonon-limited mobility of electrons and holes in wurtzite GaN using the \textit{ab initio} 
Boltzmann transport formalism, including all electron-phonon scattering processes and many-body 
quasiparticle band structures. We predict that the hole 
mobility can be increased by reversing the sign of the crystal-field splitting,
in such a way as to lift the split-off hole states above the light and heavy holes.
We find that a 2\% biaxial tensile strain
can increase the hole mobility by 230\%, up to a theoretical 
Hall mobility of 120~cm$^2$/Vs at room temperature and 620~cm$^2$/Vs at 100~K.
\end{abstract}

\maketitle

Nitride semiconductors have played a central role in the development of energy-efficient light-emitting devices (LEDs)~\cite{Ponce1997}. 
In particular, the realization of high-quality wurtzite GaN~\cite{Nakamura2015} 
enabled the successful commercialization of blue LEDs, Blu-ray optical disks, and white LED light bulbs. 
Besides its applications in photonics, GaN has been attracting considerable interest as one of the most 
promising semiconductors for power electronics~\cite{Flack2016,Amano2018}, wireless 
communications~\cite{Ishida2013}, thermoelectric energy conversion~\cite{Pantha2008}, and radiation 
detection~\cite{Atsumi2014}. Recent advances have achieved epitaxial GaN/NbN semiconductor/superconductor heterostructures~\cite{Yan2018} with the potential for quantum technologies in this material system. 
Owing to its superior electric breakdown field due to its large bandgap, and its high thermal conductivity due to the light N atoms, 
GaN based materials have potential in high voltage transistor applications for power electronics~\cite{Amano2018}. 
The two key obstacles toward this goal are the difficulty in achieving efficient p-type doping, and the low mobility of hole carriers~\cite{Amano2018}. 
While the p-type doping challenge can now be addressed via polarization-induced doping~\cite{Chaudhuri2018} and p-type field-effect transistors have  successfully been realized by this technique~\cite{Bader2018},
there is no solution in sight for improving hole mobilities, which do not exceed 40~cm$^2$/Vs at room 
temperature~\cite{Kozodoy1998,Look1999,Rubin1994,Kozodoy2000,Cheong2000,Cheong2002,Horita2017}. 
In comparison, electron mobilities as high as 1265~cm$^2$/Vs at room temperature have been reported in bulk GaN, and $>$2000~cm$^2$/Vs in 2D electron gases~\cite{Kyle2014}. 
Finding a practical strategy to increase the hole mobility of GaN would have wide-ranging implications
in wireless communications as well as quantum technologies~\cite{Yan2018}.

In this work we clarify the atomic-scale mechanisms that are responsible for the low hole mobilities
in GaN, and we propose a practical strategy for overcoming the mobility bottleneck. 
We calculate mobilities using the state-of-the-art \textit{ab initio} Boltzmann transport formalism, including all
electron-phonon scattering processes and quasiparticle GW band structures. 
The room temperature {\em electron} mobility in GaN is known to be limited by scattering from acoustic and polar optical  phonons. In contrast, we show that the origin
of the low {\em hole} mobility lies in the scattering of carriers in the light-hole ($lh$) and heavy-hole
($hh$) bands  predominantly by long-wavelength longitudinal-acoustic phonons. The heavy masses of these bands and 
the associated density of final states conspire to reduce the mobility by nearly two orders of magnitude  
compared to electron carriers. Using this understanding, our calculations predict that the hole mobility can significantly be enhanced if the split-off hole band ($sh$) can be raised above the $lh$ 
and $hh$ bands. We show that this modification of the band structure can be
achieved by reversing the sign of the crystal-field splitting via biaxial tensile strain, or 
equivalently via uniaxial compressive strain. We also show that the reversal of 
crystal-field splitting can be achieved  dynamically by coherently exciting the $A_1$ optical phonon 
via ultrafast infrared optical pulses.

The key advance that has made our present investigation possible is the recent combination of the
{\it ab initio} self-consistent Boltzmann transport equation with electron-phonon Wannier interpolation and GW quasiparticle band structures~\cite{Ponce2018}. 
This new development enables parameter-free calculations of phonon-limited mobilities in semiconductors with an unprecedented predictive power. 
In this method the carrier mobilities are computed using the linearized Boltzmann transport equation 
(BTE)~\cite{Ziman1960,Li2015,Ponce2018}, which for electrons reads:
  \begin{equation}\label{eq.1}
  \mu_{{\rm e},\alpha\beta} = \frac{-1}{n_{\rm e} \Omega} \sum_{n\in {\rm CB}}
  \int \frac{d\mathbf{k}}{\Omega_{\rm BZ}}  v_{n\mathbf{k},\alpha} \partial_{E_\beta} f_{n\mathbf{k}}.
  \end{equation}
Here $v_{n\mathbf{k},\alpha} = \hbar^{-1}\partial \varepsilon_{n\mathbf{k}}/\partial k_\alpha$ is 
the group velocity of the band state of energy $\varepsilon_{n\mathbf{k}}$, band index $n$, and wavevector $\bf k$. 
$n_{\rm e}$ is the electron density, $\partial_{E_\beta} f_{n\mathbf{k}}$ is the perturbation to the 
Fermi-Dirac distribution induced by the applied electric field $\mathbf{E}$, $\Omega$ and $\Omega_{\rm BZ}$ 
are the volumes of the crystalline unit cell and first Brillouin zone, respectively. 
Greek indices are Cartesian coordinates. 
The Berry curvature contribution to the mobility arising from the anomalous velocity vanishes
due to time-reversal symmetry, and does not appear in the BTE since we consider a homogeneous bulk system~\cite{Xiao2010}.
The perturbation to the equilibrium carrier distribution is obtained by solving the following self-consistent equation:
  \begin{multline}\label{eq.2}
  \hspace{-10pt}\partial_{E_\beta} f_{n\mathbf{k}} = e \frac{\partial f^0_{n\mathbf{k}}}{\partial 
  \varepsilon_{n\mathbf{k}}} 
  v_{n\mathbf{k},\beta} \tau_{n\mathbf{k}} + \frac{2\pi\tau_{n\mathbf{k}}}{\hbar} 
  \sum_{m\nu\sigma} \!\int\!\! \frac{d\mathbf{q}}{\Omega_{\text{BZ}}} | g_{mn\nu}(\mathbf{k,q})|^2 \\
  \hspace{-10pt}\times \![ (1\!+\!\sigma)/2\!-\! \sigma f_{n\mathbf{k}}^0 \!+\! n_{\mathbf{q}\nu} ] 
  \delta( \Delta \varepsilon^{mn}_{\mathbf{k},\mathbf{q}} \!+\! \sigma\hbar \omega_{\mathbf{q}\nu} )  
  \partial_{E_\beta} f_{m\mathbf{k+q}},\hspace{-5pt}
  \end{multline}
where $\sigma \pm 1$ stands for phonon absorption/emission, $\Delta \varepsilon^{mn}_{\mathbf{k},\mathbf{q}} =
\varepsilon_{n\mathbf{k}} - \varepsilon_{m\mathbf{k+q}}$, and $f^0_{n\mathbf{k}}$ is the equilibrium 
distribution function. 
The matrix elements $g_{mn\nu}(\mathbf{k}, \mathbf{q})$ are the probability 
amplitudes for scattering from an initial electronic state $n{\bf k}$ to a final state $m{\bf k+q}$
via a phonon of branch index $\nu$, crystal momentum $\bf q$, and frequency $\omega_{\mathbf{q}\nu}$.
$n_{\mathbf{q}\nu}$ is the Bose-Einstein occupation of this mode.  
The quantity $\tau_{n\mathbf{k}}$ in Eq.~\eqref{eq.2} is the relaxation time, and is given by~\cite{Grimvall1981,Giustino2017}:
  \begin{multline}\label{eq.3}
  \frac{1}{\tau_{n\mathbf{k}}} = \frac{2\pi}{\hbar} \sum_{m\nu\sigma}\int \frac{d\mathbf{q}}{\Omega_{\rm BZ}}  
  |g_{mn\nu}(\mathbf{k},\mathbf{q})|^2  \\ \times [(1+\sigma)/2 - \sigma f_{m\mathbf{k}+\mathbf{q}}^0 
  + n_{\mathbf{q}\nu}] 
  \delta(\Delta \varepsilon^{mn}_{\mathbf{k},\mathbf{q}} - \sigma \hbar\omega_{\mathbf{q}\nu}).
  \end{multline}
In our calculations we first compute Eq.~\eqref{eq.3}, then solve Eq.~\eqref{eq.2} iteratively to
obtain $\partial_{E_\beta} f_{n\mathbf{k}}$, and we use the result inside Eq.~\eqref{eq.1}.
Since most experiments report Hall-effect mobilities instead of drift mobilities, we also compute the
Hall factor and scale our results accordingly, as discussed in the accompanying manuscript~\cite{Ponce2019_PRB}.
All calculations are based on density-functional theory including spin-orbit coupling (SOC) for the 
Kohn-Sham states, density-functional perturbation theory for phonons and electron-phonon matrix elements, 
and many-body perturbation theory for GW quasiparticle corrections, as implemented in the software 
packages Quantum Espresso~\cite{Giannozzi2017}, Yambo~\cite{Marini2009}, wannier90~\cite{Mostofi2014}, and EPW~\cite{Giustino2007,Ponce2016a}. 
Calculation details are provided in Ref.~\onlinecite{Ponce2019_PRB}.

\begin{figure}[ht]
  \centering
  \includegraphics[width=0.98\linewidth]{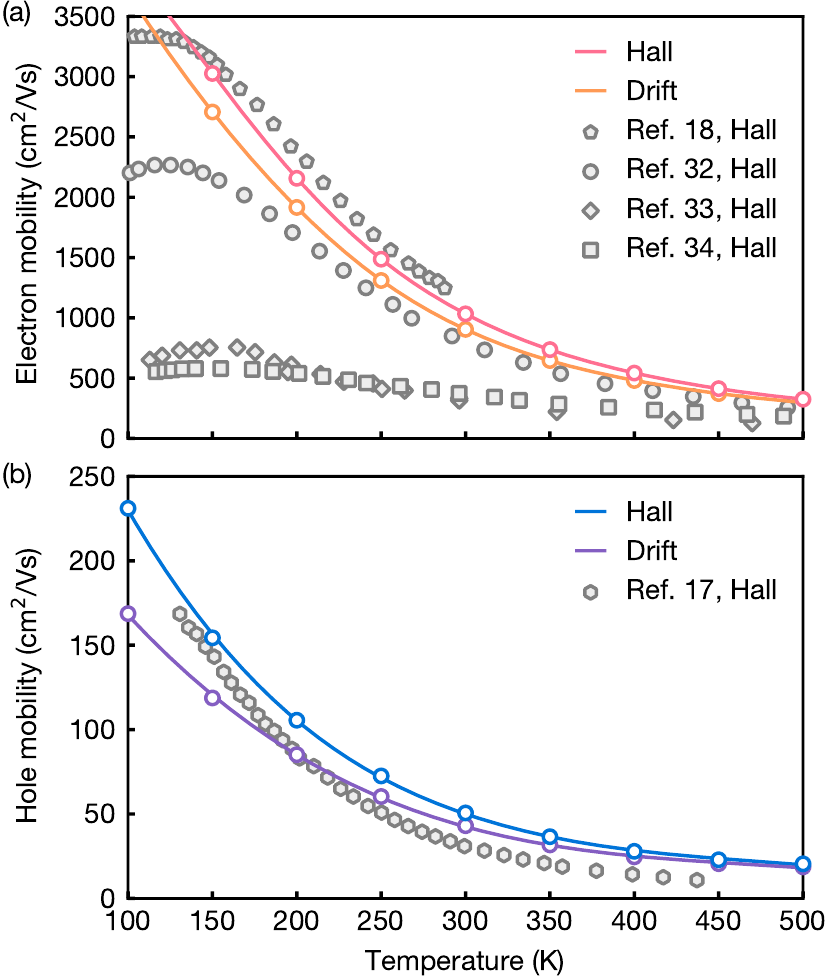}
  \caption{\label{fig1} 
  Electron (a) and hole (b) mobilities of wurtzite GaN, calculated using the \textit{ab initio} Boltzmann
  formalism, compared with the experimental data from 
  Ref.~\onlinecite{Kyle2014,Goetz1998,Ilegems1973,Goetz1996,Horita2017}. We show both the drift mobilities computed via Eqs.~(\ref{eq.1})-(\ref{eq.3}) and the Hall 
  mobilities obtained by applying the Hall factor taken from Ref.~\onlinecite{Ponce2019_PRB}.
  }
\end{figure}

In Fig.~\ref{fig1} we show our calculated drift and Hall-effect mobilities for electrons and holes in
intrinsic GaN as a function of temperature, and we compare our results with available experimental data. 
We find that the Hall mobility is about 15\% higher than the drift mobility at room temperature, as expected~\cite{Lundstrom2009}.
Our predicted electron Hall mobility of 1030~cm$^2$/Vs and hole Hall mobility of 50~cm$^2$/Vs at 300~K are in good agreement with the measured values  
1265~cm$^2$/Vs~\cite{Kyle2014} and 31~cm$^2$/Vs~\cite{Horita2017}, respectively.

In order to clarify the origin of the large difference between the electron and hole mobilities in GaN we refer to Eqs.~\eqref{eq.1}-\eqref{eq.3}. 
In the simplest case of parabolic bands,  in line with Drude's law, the mobility in Eq.~\eqref{eq.1} scales as 
$e \tau/ m^*$, with $m^*$ and $\tau$ average effective mass and relaxation rate, respectively. 
As detailed in the accompanying manuscript~\cite{Ponce2019_PRB}, the conduction band bottom of GaN is singly degenerate, while 
at the valence band top we have a $lh$ and a $hh$, which are split into doublets by SOC as we move from the $\Gamma$ to M points of the Brillouin zone. 
The split-off hole will be discussed shortly.
Our calculated effective masses are $m_{lh}^{\perp/\parallel} = 0.37/1.66\,m_{\rm e}$, 
$m_{hh}^{\perp/\parallel} = 0.45/1.94\,m_{\rm e}$, and $m_{e}^{\perp/\parallel} = 0.20/0.23\,m_{\rm e}$, 
where $\perp$ and $\parallel$ denote the in-plane $\Gamma$%
$\rightarrow$M or $\Gamma$%
$\rightarrow$K directions, 
and the out of plane $\Gamma$%
$\rightarrow$A directions in the Brillouin zone. 
These values are in reasonable agreement with experimental data ranging from $0.3\,m_{\rm e}$ to $2.03\,m_{\rm e}$~\cite{Pankove1975,Xu1993,Fan1996,Yeo1998,Rodina2001} for holes 
and in good agreement with $0.2\,m_{\rm e}$~\cite{Drechsler1995} for the electrons. 
%
%
The ratio between the conductivity effective masses (i.e. the harmonic averages) of electrons and holes is 2.4-2.9 for the $hh$/$lh$ case. 
These values are much smaller than the ratio between the calculated mobilities at room temperature ($1030/50\approx 21$), therefore the difference between electron 
and hole masses alone cannot fully account for the order-of-magnitude difference in carrier mobilities.

\begin{figure}[t]
  \centering
  \includegraphics[width=0.98\linewidth]{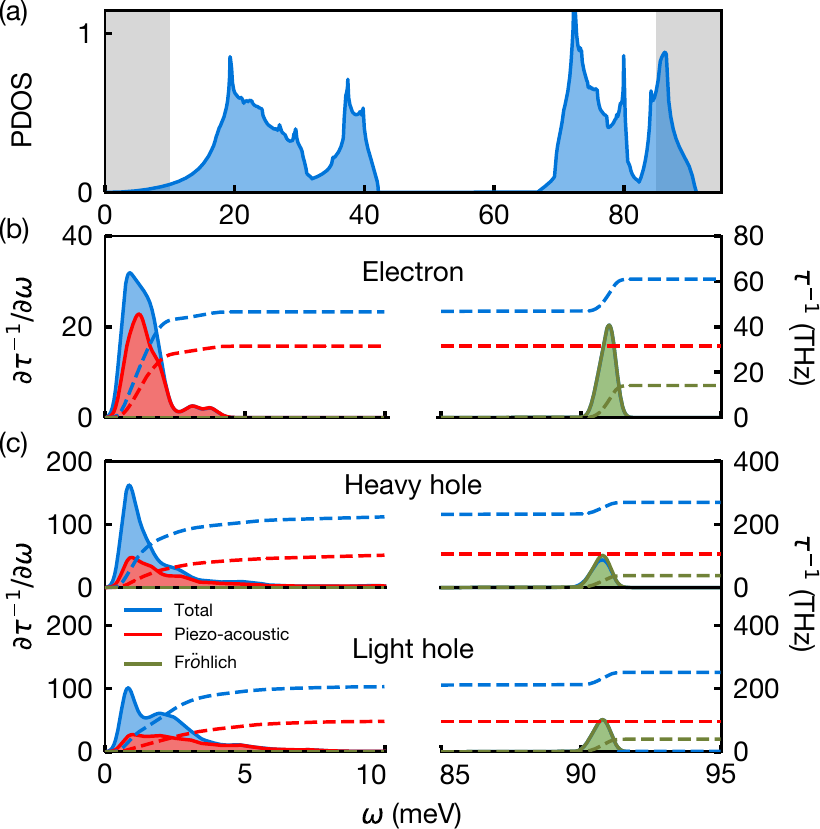}
  \caption{\label{fig2}  
  Phonon density of states in GaN (a), and spectral decomposition of the electron (b) and hole (c) 
  scattering rates as a function of phonon energy where the piezo-acoustic (red) and polar Fr\"ohlich (green) have been highlighted. 
  The rates are calculated as angular averages for carriers at an energy
  of $k_{\rm B}T=25$~meV away from the band edge (left vertical axis). The dashed lines 
  represent the cumulative integral of each spectrum of $\partial \tau^{-1}/\partial\omega$, 
  and add up to the carrier scattering rate $\tau^{-1}$ (right vertical axis). 
  The shadings in (a) indicate the energy ranges shown in (b) and (c).
  }
\end{figure}

To identify the origin of the residual difference between electron and hole mobilities, in Fig.~\ref{fig2} we show the carrier relaxation rate $1/\tau$. 
Although every electronic state has its own lifetime $\tau_{n\mathbf{k}}$, for definiteness we report the values for carriers at an energy 
$k_{\rm B}T = 25$~meV away from the band edges. 
We have shown previously that this is the most representative carrier energy~\cite{Ponce2019a}. 
Alongside the total relaxation rate, in this figure we also show the spectral decomposition of $1/\tau$ in terms of phonon energy and the
phonon density of states for comparison. 
The main scattering channel is from long-wavelength acoustic phonons [peaks around a phonon energy of 2~meV in 
Fig.~\ref{fig2}(b) and (c)], which account for 77\% and 84\% of the scattering rates for electrons and holes, respectively. 
The remaining contribution is from the longitudinal-optical (LO) phonon near $\Gamma$, which is responsible for Fr\"ohlich electron-phonon coupling. 
The scattering due to acoustic phonons can be further broken down into a piezo-acoustic and an acoustic-deformation-potential (ADP) contribution~\cite{Lundstrom2009}. 
The latter accounts for 48\% and 61\% of the total scattering for electrons and holes, respectively. 

In the case of ADP scattering, the rates in Eq.~\eqref{eq.3} can be approximated by neglecting the phonon 
frequency in the Dirac delta function, and taking the matrix elements as a constant. 
This leads to a scattering rate that scales with the electronic density of states, and hence with the effective masses, 
as $1/\tau \sim m^{*,3/2}$. Our data approximately follows this trend. 
In fact the electron lifetimes in Fig.~\ref{fig2} are in the range of 17~fs, while the hole lifetimes are much shorter, 
around 4~fs. Their ratio  $17/4\approx 4$ is of the same magnitude as the corresponding ratio 
between the conductivity effective masses, $m_{lh}^{*,3/2}/m_{\rm e}^{*,3/2} = 3.7$ and $m_{hh}^{*,3/2}/m_{\rm e}^{*,3/2} = 4.9$. 
This finding indicates that the high density of states associated with the $lh$ and $hh$ bands plays a central role in reducing hole lifetimes and hence 
suppressing their mobility. 
Other effects also contribute to reducing hole mobilities, albeit to a lesser extent; 
for example the presence of multiple scattering channels for the holes (two spin-split sets of 
bands), the strong non-parabolicity of the $hh$ band~\cite{Kim1996, Yeo1998,Rinke2008,Svane2010}, and the longitudinal-optical phonons. 
All these effects are fully accounted for in our \textit{ab initio} BTE formalism. 
We note that this analysis holds for electron or hole states $\sim$25~meV from the band edges,
and should be re-evaluated for heavily degenerate 2D electron or hole gases.

\begin{figure*}[ht]
  \centering
  \includegraphics[width=0.99\linewidth]{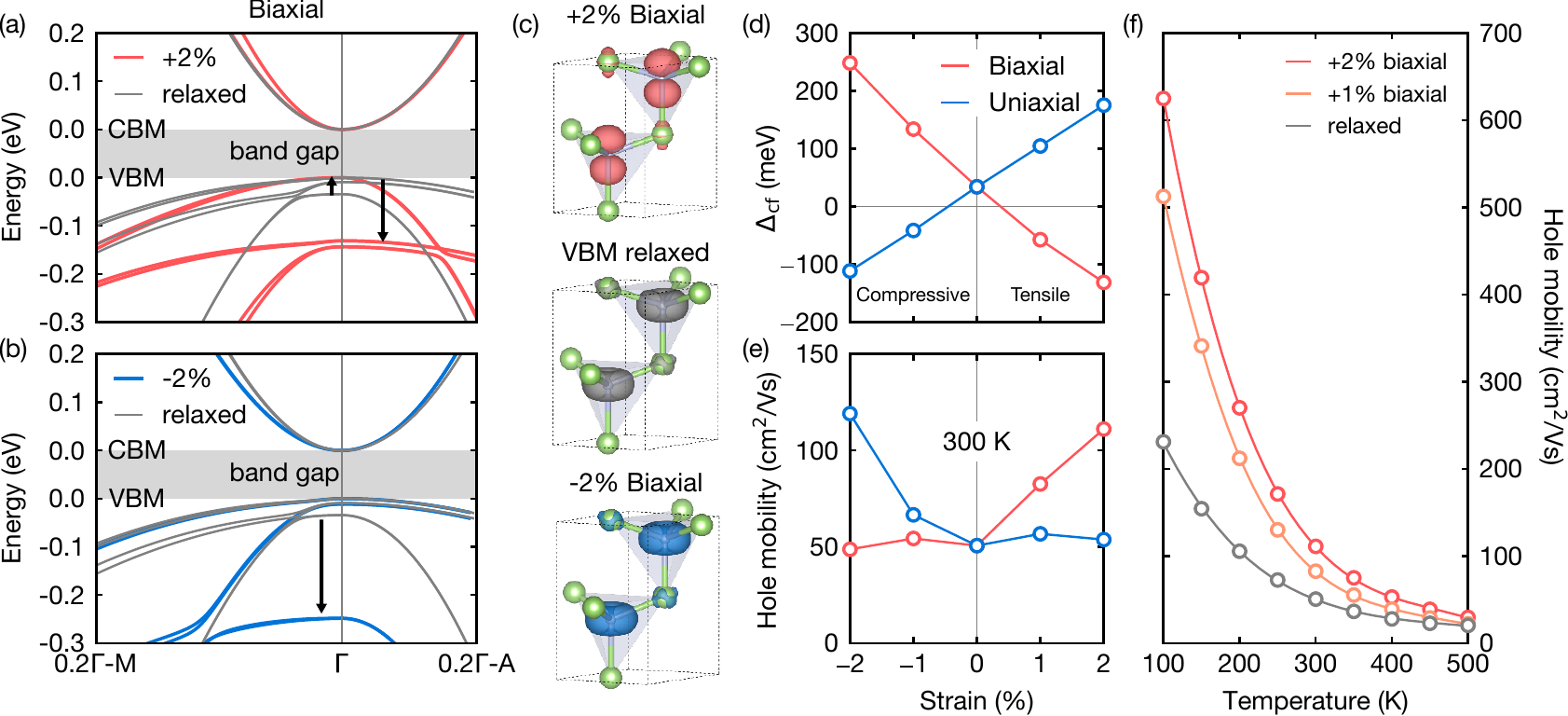}
  \caption{\label{fig3} 
  Crystal-field engineering of band structure and mobility in GaN. (a), (b) Change in the GW quasiparticle
  band structure of GaN upon biaxial dilation and compression, respectively. The energy levels have been aligned to the conduction and valence band edges. (c) Electron wavefunction
  at the valence band maximum at $\Gamma$ for the undistorted wurtzite GaN structure, as well as for 
  2\% biaxial dilation and 2\% biaxial compression, respectively. 
  (d)~Crystal-field splitting $\Delta_{\rm cf}$ vs.\ strain, and (e) corresponding hole Hall mobility at 300~K. (f) Predicted 
  temperature-dependent hole mobility in wurtzite GaN as a function of biaxial strain.
  }
\end{figure*}

How can one improve the hole mobility in GaN?
Since the low mobilities stem primarily from the presence of two adjacent bands with heavy masses, one possibility is to check whether we can remove 
at least one of these bands via strain engineering, or alternatively bring the light $sh$ hole nearer to the valence band top. 
It turns out that the $lh$ and $hh$ bands are separated by the spin-orbit splitting $\Delta_\text{so}$, 
and this splitting is relatively insensitive to strain as discussed in our accompanying manuscript~\cite{Ponce2019_PRB}. 
However, the separation between the $lh$/$hh$ and the $sh$ bands is controlled by the crystal-field splitting, $\Delta_{\text{cf}}$.
This quantity is known to be sensitive to the internal parameter $u$ of the wurtzite structure, 
or equivalently to a change of the $c/a$ ratio~\cite{Kim1996,Yan2009}. 
In Fig.~\ref{fig3}(d) we show that $\Delta_\text{cf}$ can indeed be altered by biaxial or 
uniaxial strain. Most importantly, a lattice distortion that reduces the $c/a$ ratio or increases the $u$ parameter can \textit{reverse the sign}
of the crystal-field splitting; this is the case for biaxial tensile strain (along [2$\overline{1}\overline{1}$0] and [$\overline{1}$2$\overline{1}$0])
and for uniaxial compressive strain (along [0001]). 
Under these conditions we can have the $sh$ band being lifted above the $lh$ and $hh$ bands, 
as shown in Fig.~\ref{fig3}(a). This effect alters the ordering of the valence band top, as well as the character of the wavefunctions, see Fig.~\ref{fig3}(c). 
In particular the hole wavefunctions go from N-$p_{x,y}$ states to N-$p_z$ states. 
If we consider instead a biaxial compressive strain or uniaxial tensile strain, the $c/a$ ratio increases and $\Delta_\text{cf}$ maintains the same sign but increases 
in magnitude. 
In this case the band ordering remains unchanged from the situation in bulk unstrained GaN. 
The conductivity effective mass of the $sh$ band is $m^*_{sh} = 0.45\,m_{\rm e}$ at the 
zone center, but away from $\Gamma$ it quickly decreases to $m^*_{sh} = 0.22\,m_{\rm e}$ due to strong 
non-parabolicity. Since this value is much smaller than the masses of the $lh$ and $hh$ bands, 
we hypothesize that the hole mobility of GaN will improve upon reversal of the sign of $\Delta_\text{cf}$.

To test our hypothesis we repeat all transport calculations for strained GaN.
Figure~\ref{fig3}(f) shows the hole mobility of GaN computed for biaxial tensile strains of 1\% and of 2\%. 
Similar results can be found for the case of uniaxial compressive strain~\cite{Ponce2019_PRB}. 
The electron mobility is much less affected by strain.
The results confirm indeed our expectation that, as soon as we change the sign of $\Delta_\text{cf}$, we have an 
enhancement in the hole mobility. In particular, the comparison between Fig.~\ref{fig3}(d) and Fig.~\ref{fig3}(e) shows that 
the hole mobility nicely tracks the quantity $\max(-\Delta_\text{cf},0)$, thus further confirming our point.

As shown on Fig.~\ref{fig3}(f) and in the companion manuscript~\cite{Ponce2019_PRB}, we predict a significant increase of the hole 
mobility at room temperature, up to $\sim$120~cm$^2$/Vs under 2\% biaxial dilatation or 2\% uniaxial compression.
This value represents a 230\% increase in the hole mobility with respect 
to unstrained GaN. We emphasize that our results are not sensitive to the details of the calculations, 
since they are rooted in a \textit{change of band character} of the valence band top under an elastic deformation. 
This is confirmed by explicit calculations of $\Delta_\text{cf}$ using different functionals.

How realistic is our proposal? 
We computed the GaN phase diagram to check that the wurzite structure remains the lowest-enthalpy phase under strain~\cite{Ponce2019_PRB}.
Engineering biaxial strain of up to 4\% is feasible nowadays by growing GaN on epitaxial substrates such as AlN or 6H-SiC~\cite{Jain2000,Wagner2002}. 
Furthermore the most common substrate for growing GaN is 6H-SiC, and the lattice mismatch in this case is 3.5\%~\cite{Jain2000}. 
The reason why high hole mobilities have not yet been observed for GaN grown on these substrates may be that GaN films are so thick that the crystal 
lattice accommodates the strain via misfit dislocations~\cite{Floro2004}, 
which further decrease the mobility. Therefore to realize high-hole-mobility GaN one should devise strategies for preventing dislocation nucleation. 

One possibility would be to grow GaN films thinner than the Fischer critical thickness for misfit dislocations~\cite{Matthews1974,Fischer1994} which we computed to be 7~nm at 2\% strain (see accompanying manuscript~\cite{Ponce2019_PRB}) and should also mitigate cracking under stress~\cite{Dreyer2015}. 
For GaN on AlN it was recently shown that the critical thickness ranges between 3 and 30~monolayers depending on the growth temperature~\cite{Sohi2017}. 
Such layer thicknesses have recently become possible~\cite{Qi2017,Islam2017}, therefore our present proposal should be feasible
by carefully controlling the growth conditions. 
The switching of the valence band ordering could be detected optically by measuring the polarization of optical emission from such layers~\cite{Park2011}.
%
%
Furthermore in previous work the GaN samples were thicker than the Matthews-Blakeslee critical thickness~\cite{Matthews1974}, therefore in those cases the hole mobility enhancement is expected to be suppressed by dislocation scattering.

Another intriguing possibility for increasing the hole mobility of GaN could be to tune the crystal-field splitting via the internal parameter $u$.
Since this parameter is controlled by the $A_1$ transverse-optical phonon at $\Gamma$, it should be possible to control the hole mobility via light, by coherently exciting optical 
phonons with femtosecond infrared pulses. 
Light-induced control of transport coefficients has already been demonstrated for NdNiO$_3$~\cite{Caviglia2012}, and recent developments in the 
parametric amplification of optical phonons~\cite{Cartella2018} offer many new opportunities in this direction.

In summary, we have shown that the origin of the low hole mobilities in GaN lies in a combination of heavy carrier effective masses and high density of final electronic states available 
for scattering via low-energy acoustic phonons. 
We find that this bottleneck can be circumvented by modifying the ordering of the valence band top, in such as way as to have the split-off holes 
above the light holes and heavy holes. 
We propose to realize such band inversion by reversing the sign of the crystal-field splitting via strain engineering or via optical 
phonon pumping. We hope that this work will stimulate experimental research in high-hole-mobility 
GaN and will accelerate progress towards GaN-based CMOS technology.

\begin{acknowledgments}
We are grateful to E.~R.~Margine for assistance with the calculation of the band velocity,
and M.~Schlipf for useful discussions. This work was supported by the Leverhulme Trust 
(Grant RL-2012-001), the UK Engineering and Physical Sciences Research Council (grant 
No. EP/M020517/1), the Graphene Flagship (Horizon 2020 Grant No. 785219 - GrapheneCore2),
the University of Oxford Advanced Research Computing (ARC) facility 
(http://dx.doi.org/810.5281/zenodo.22558), the ARCHER UK National Supercomputing Service under 
the AMSEC and CTOA projects, PRACE DECI-13 resource Cartesius at SURFsara, the PRACE DECI-14 resource 
Abel at UiO, and the PRACE-15 and PRACE-17 resources MareNostrum at BSC-CNS. 
DJ acknowledges support in part from the NSF DMREF award \# 1534303 monitored by Dr. J. Schluter, NSF Award \# 1710298 monitored by Dr. T. Paskova, the NSF CCMR MRSEC Award \#1719875,  AFOSR under Grant
FA9550-17-1-0048 monitored by Dr. K. Goretta, and a research grant from Intel.
\end{acknowledgments}

\end{document}